\documentclass{rmf-d}
\usepackage{nopageno,rmfbib,multicol,times,epsf,amsmath,amssymb,cite}
\usepackage[latin1]{inputenc}
\usepackage[]{caption2}
\usepackage{graphicx}
\usepackage{float}
\usepackage{baskervald}
\def\rmfcornisa{}
\def\rmfcintilla{}
\clearpage \rmfcaptionstyle 
\begin{document}
\title{Physics of the tau lepton
\vspace{-6pt}}
\author{Jorge Portol\'es \footnote{Email: Jorge.Portoles@ific.uv.es}} 
\address{Instituto de F\'{\i}sica Corpuscular, CSIC - Universitat de Val\`encia, \\ Apt. Correus 22085, E-46071 Val\`encia, Spain}
\maketitle
\begin{abstract}
\vspace{1em} Within our present knowledge, the tau is the heaviest lepton and the only one decaying into hadrons, a fact that makes it the source of 
a very rich phenomenology. It represents the third family of leptons in the Standard Model, a feature that helps its classification but whose real meaning
is not asserted yet. The tau lepton provides: i) a clean and unique environment to study  both the hadronization of QCD currents, in an energy region populated by 
resonances, and the phenomenological determination of relevant parameters of the Model; ii) together with the muon, they have a very constrained flavour dynamics (in the absence of neutrino masses) due to an accidental global symmetry of the 
Standard Model. In consequence, the tau lepton brings an excellent benchmark for the study of QCD at low energies and, at the same time,  for the search of new physics.

\vspace{1em}
\end{abstract}
\keys{Phenomenology of the tau lepton \vspace{-4pt}}
\pacs{ \bf{\textit{14.60.Fg, 13.35.-r}}    \vspace{-4pt}}
\begin{multicols}{2}
\section{Introduction}
\label{sec:intro}
The Standard Model of particle physics (SM) \cite{Donoghue:1992dd,Pich:2012sx,Schwartz:2014sze} is defined by a fundamental local $SU(3)_C \times SU(2)_L \times U(1)_Y$ gauge symmetry, a Higgs field and a well defined spectrum of matter that includes quarks and leptons, that interact quirally, with the electroweak gauge bosons and Higgs, i.e. through $SU(2)$ left doublets and right singlets. A part of the fact that leptons have no color and do not feel the strong force, there are some important differences in the electroweak structure of quarks and leptons: while the quark families are constituted by two flavours, lepton families are made of a neutrino and a charged lepton, both with the same flavour, and there is no right-handed neutrino. 
Moreover, for reasons that do not seem related with the fundamental symmetry, it is kind of accidental that there are three families of quarks and leptons. 
A relevant feature to point out is that while the three quark families correspond to six flavours, the labels of flavour, generation or family are 
equivalent for leptons: we have three, namely, $e$, $\mu$ and $\tau$.  In addition, while we have a very rich quark flavour dynamics, lepton flavour is conserved in all processes. In this note we will dwell on the physics of the tau lepton, the third family. Third because the orders of discovery and increasing mass.  
\par 
The discovery of the muon lepton in cosmic-ray showers \cite{Neddermeyer:1937md} produced a question about the differences between electron and muon. Apart of 
their different masses it did not seem that they had any other distinction. In the early seventies of the twentieth century the question  was still around and 
prevailed the general mood that heavier leptons could also exist and could be observed with the new colliders \cite{Tsai:1971vv}. Then, in 1975 the collaboration of the Mark I detector at the $e^+e^-$ collider in SLAC, sifting through 35000 events, found 24 with a $\mu$ corresponding to an opposite sign $e$, i.e. $e^+ e^- \rightarrow \mu^{\pm} e^{\mp}$ \cite{Perl:1975bf}. These anomalous \lq \lq$\mu \, e$" events represented a puzzle that could be explained by the creation and decay of a couple
of heavy leptons, the tau leptons to be, namely: $e^+ e^- \rightarrow \tau^+ \tau^-$ with the decays $\tau^+ \rightarrow e^+ \nu_e \overline{\nu}_{\tau}$ and
$\tau^- \rightarrow \mu^- \overline{\nu}_{\mu} \nu_{\tau}$ , with a mass $M_{\tau} \sim 2 \, \mbox{GeV}$ and hypothesizing the existence of a new tau neutrino, $\nu_{\tau}$. A later confirmation came in 1977 from the PLUTO collaboration at the DORIS $e^+ e^-$ storage ring \cite{PLUTO:1977ctk}, and finally the conclusion that the dynamics of the tau lepton in the SM was the same than electron and muon was asserted by the ARGUS detector at the DORIS II storage ring \cite{ARGUS:1990jan} in 1990. The direct observation of the tau neutrino took place ten years later by the DONuT collaboration at Fermilab \cite{DONUT:2000fbd}. 
\par 
Although the dynamics of the tau lepton has thoroughly been studied since its discovery, and some experiments have contributed to its phenomenological
analysis, it has been the start of the 21st century that has pushed the physics of the tau lepton with the development of the B-meson factories: BaBar at SLAC (1999-2008) \cite{BaBar:1998yfb} and Belle at KEK (1999-2010) \cite{Belle:2000cnh,Miyabayashi:2001uh}. These are asymmetric $e^+ e^-$ colliders producing plenty
of B mesons but they happen to be tau lepton factories too. Although their data acquisition period has ended they still have enough data to be analysed. The present Belle-II experiment at SuperKEK (an upgrade of Belle) has started to collect data in 2019 \cite{Belle-II:2018jsg} and, with an expected integrated luminosity of 50 $\mbox{ab}^{-1}$, will push the frontier of our phenomenological analyses of tau decays.
\par 
Within the SM the setting provided by the tau lepton is unique. As the only known lepton to be heavier enough to decay into light flavoured hadrons, it brings a
benchmark for the studies that involve strong interactions at low energies and the dynamics of hadronization. The same basic reason is behind the accurate determination
of some SM parameters. This goal has guided a big part of the amount of work done on the tau lepton.  Besides, in the last ten years the tau lepton has been at 
the origin of some seeming deviations of the universality SM rule, that says that, for massless neutrinos, all leptons of equal electric charge have the same electroweak interactions, independently of their flavour. Departures of this principle have been reported by the LHCb experiment, at LHC, in semilepton decays of B mesons, although as of today, there is no asserted discovery of new physics \cite{London:2021lfn}.
\par 
Although there is a very rich phenomenology around the tau lepton in many processes, in this text I will only focus on the features that involve its decays. 
In Section~2 I will recall some basic properties of the tau lepton and relevant aspects of its dynamics in the SM. The analyses of tau decays within the SM,
both lepton and hadron, will be collected in Section~3. In Section~4 I will provide a quick look to the issue of lepton flavour violation as a promise of 
new physics in the tau sector. My conclusions and summary are collected in Section~5. 
\section{Dynamics and properties of the tau lepton}
\label{sec:DYN}
The tau lepton has two properties that mark the difference with the rest of leptons. One of them is its high mass, in comparison with $e$ and $\mu$, \cite{ParticleDataGroup:2020ssz}
\begin{equation}  \label{eq:tauM}
M_{\tau} \, = \, 1.77686 \,(12) \, \mbox{GeV} \, , 
\end{equation}
where the number in parentheses indicates the error of the last corresponding figures. As a consequence, SM dynamics allowing, it becomes the only known lepton that can decay into light-flavoured hadrons. The second property is related with the global symmetries of the SM lagrangian \cite{Chivukula:1987py}. In the presence of Yukawa couplings but with massless neutrinos, it has a global symmetry
\begin{equation} \label{eq:acci}
U(1)_e \times U(1)_\mu \times U(1)_\tau \times U(1)_B \times U(1)_Y \, ,
\end{equation} 
where $B$ is short for baryon number and $Y$ is the weak hypercharge. This symmetry has relevant consequences: i) different leptons are characterized by a specific flavour that is conserved in all processes in the SM (with massless neutrinos); another consequence is that lepton number, $L=N_e+N_{\mu}+N_{\tau}$, is also conserved; ii) baryon number is conserved in all SM processes. The later feature brings more information on the hadron decays of the tau: although there is enough phase space to decay into baryons  (proton, $\Lambda$, $\Sigma$, ...), there is no
enough phase space for a pair of them and, accordingly, the tau lepton cannot decay into baryons, only mesons are allowed. 
\par
\begin{figure}[h]
\centering
     \includegraphics[scale=0.7]{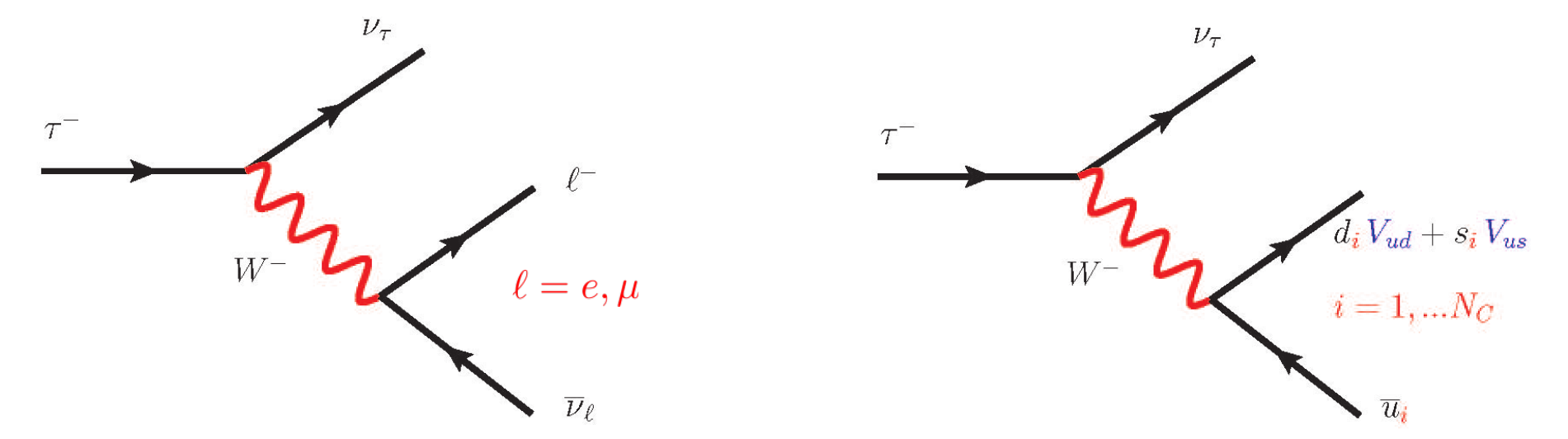}%
\caption{\label{fig:1} Leading tree-level Feynman diagrams describing the decays of the tau lepton in the SM, for lepton final states (left) and hadron final states (right). Here $V_{ud}$ and $V_{us}$ are the CKM matrix elements. }
\end{figure}

 
In the SM tau decays are driven by the charged current of leptons
\begin{equation} \label{eq:chargedc}
{\cal L}_{\mbox{\tiny SM}}  \supset  - \sum_{i = e, \mu, \tau} \frac{g_i}{2 \sqrt{2}}  \left[ \overline{\nu}_i \, \gamma^{\mu} \, (1-\gamma_5) \, \ell_i \, W_{\mu}^{\dagger} \, \right] +  h.c. \, , 
\end{equation}
with $g_e = g_\mu = g_\tau = g$, the $SU(2)_L$ coupling. This current drives the tau decay into leptons $\tau^- \rightarrow \nu_{\tau} \ell^- \overline{\nu}_{\ell}$,
for $\ell =e,\mu$ and those with final quarks $\tau^- \rightarrow \nu_{\tau} (\overline{u} d, \overline{u} s)$ and charge conjugates.
Only with this information and the corresponding one for hadrons, as shown in Fig.~\ref{fig:1}, we 
can make good estimates for the exclusive branching ratios into leptons and the inclusive decays into hadrons. Notice that the total number of decays comes from the possible 2 lepton final states added to those into hadrons, i.e. for the $d$ quark a width of $|V_{ud}|^2$ times the number of possible final quarks, i.e $N_C$, the  quark colours, and analogously for the $s$ quark, for a total of $2 + N_C(|V_{ud}|^2 + |V_{us}|^2)$ full width. We show the figures in Table~\ref{tab:1}; the agreement is fairly good for this rough guess. The interesting fact is that more accurate determinations within the SM are able to correct these naive estimates and explain reasonably well the experimental measurements. 
\end{multicols}
%
\begin{table}[h]
\begin{center}
\renewcommand{\arraystretch}{1.5}
\begin{tabular}{|c|c|c|}
\hline
Branching & Estimate  & Experiment \cite{ParticleDataGroup:2020ssz} \\
\hline
\hline
$\mbox{Br}(\tau \rightarrow e \overline{\nu} \nu)$ & $\left[ 2 + N_C \left( |V_{ud}|^2 + |V_{us}|^2 \right) \right]^{-1}$ 
& $17.82 (4) \%$ \\
$\mbox{Br}(\tau \rightarrow \mu \overline{\nu} \nu)$ & $ \simeq \, 20 \%$ 
& $17.39 (4) \%$ \\
\hline
$\mbox{Br}(\tau \rightarrow \mbox{non-strange hadrons})$ & $N_C |V_{ud}|^2 \, \left[ 2 + N_C \left( |V_{ud}|^2 + |V_{us}|^2 \right) \right]^{-1} \, \simeq \, 
58 \%$ & $ 62 (4) \%$ \\ 
\hline
$\mbox{Br}(\tau \rightarrow \mbox{strange hadrons})$ & $N_C |V_{us}|^2 \, \left[ 2 + N_C \left( |V_{ud}|^2 + |V_{us}|^2 \right) \right]^{-1} \, \simeq \, 
2 \%$ & $ 2.6 (7) \%$ \\ 
\hline
\end{tabular}
\end{center}
\caption{\label{tab:1} Comparison between a naive estimate and its experimental result for some tau decay branching ratios, as determined by the Feynman diagrams in Fig.~\ref{fig:1}.  } 
\end{table}
\begin{multicols}{2}
\section{Tau decays within the Standard Model}
\label{sec:SM}
We will now dwell on the richness and variety of the tau decay processes driven by the diagrams in Fig.~\ref{fig:1}. I will only bear in mind, in this note, dominating
processes and I will not take into consideration subleading radiative processes: photons can be attached where any electrically charged particle lies. In the lepton decays, to the natural cleanliness of the processes we add, for the first time, two possible decay channels; this situation will allow us to know more about the
flavour aspects of these leptons. Moreover, in hadron decays, being an initial decaying lepton, the produced hadrons decouple from the initial state and it is driven by the hadronization of the charged current in low-energy QCD. Along this note we will consider the same dynamics for charge conjugated processes and their equivalence will be understood. We now underline the main physics features of both decay types in turn.
\subsection{Lepton decays}
\label{subsec:LSM}
The left-hand diagram in Fig.~\ref{fig:1} gives the leading contribution for the decays $\tau^- \rightarrow \nu_{\tau} \ell^- \overline{\nu}_{\ell}$ for 
$\ell = e,\mu$. Once the dominant electroweak corrections are included, the width of the decay is \cite{Marciano:1988vm}
\begin{equation} \label{eq:leptoms}
\Gamma(\tau^- \rightarrow \nu_{\tau} \ell^- \overline{\nu}_{\ell})  =  \frac{G_F^2 \, M_{\tau}^5}{192 \pi^3} \, f(M_{\ell}^2/M_{\tau}^2) \, r_{\mbox{\tiny EW}} \, ,  
\end{equation}
where the higher order electroweak correction is given by 
\begin{equation} \label{eq:rew}
r_{\mbox{\tiny EW}} = \left(1+\frac{3}{5} \frac{M_{\tau}^2}{M_W^2} \right) \left[ 1 + \frac{\alpha(M_{\tau})}{2 \pi} \left( \frac{25}{4} - \pi^2 \right) \right]  \, 
\end{equation}
that amounts to $r_{\mbox{\tiny EW}} \simeq 0.9960$. In Eq.~(\ref{eq:leptoms}), $G_F$ is the Fermi constant and the corrections due to the mass of the final lepton are encoded in $f(x) = 1 - 8x + 8x^3-x^4 - 12x^2 \ln x$ that are tiny $f(M_e^2/M_{\tau}^2) \simeq 0.999999$ or very small $f(M_\mu^2/M_{\tau}^2) \simeq 0.972559$. As a 
consequence the SM width, dominated by the first factor on the right-hand side of Eq.~(\ref{eq:leptoms}), is almost independent of the final lepton, as our rough guess
and the experimental measurements already were pointing out in Table~\ref{tab:1}. 
\par 
This scenario is a result of the equality of couplings in the lagrangian (\ref{eq:chargedc}), a feature of the SM known as universality of the lepton couplings, that is
spoiled for massive neutrinos.  The cleanest way to study this universality involves the decays of the gauge boson, i.e. $W^- \rightarrow \ell^- \overline{\nu}_{\ell}$
for $\ell = e,\mu,\tau$. If one takes a look to the ratios of widths, that the SM predicts to be 1, in the PDG \cite{ParticleDataGroup:2020ssz} we have:
\begin{eqnarray} \label{eq:ratioW}
\mu / e & = & 0.996 (8) \, , \nonumber \\
\tau / e & = & 1.043 (24) \, ,  \\
\tau / \mu & = & 1.070 (26) \, \nonumber
\end{eqnarray}
in a self-explanatory notation. These results come from old LEP data  and show a tension related with the tau coupling. The possibility of a breaking of universality
centered in the third family, i.e. imposing a global symmetry  $[U(2)_{e,\mu} \times U(1)_{\tau}]^5$ that distinguishes the $g_\tau$ coupling from the one of electron
and muon (\ref{eq:chargedc}), coming from an energy scale much above the electroweak one, was analysed in Refs.~\cite{Han:2005pr,Filipuzzi:2012mg} with no avail. This
breaking could not explain the seeming violation of universality. However the ATLAS collaboration at LHC \cite{ATLAS:2021icw} recently provided a new result
\begin{equation} \label{eq:buniv}
\tau / \mu = 0.992 (13) \, , 
\end{equation} 
in good agreement with the universality principle.  This shows that more precise experimental results are required to settle this issue. 
\par 
The dynamical structure of the coupling of the leptons to the gauge boson in the charged current (\ref{eq:chargedc}) is predicted to be V-A in the SM. We already
know that this feature is well established but possible deviations from the SM predictions could be asserted at the B-factories. This goal can be achieved through
the Michel parameters \cite{Michel:1949qe,Rouge:2000um} $g_{ij}^a$, in general complex, defined by the matrix element of the tau decay
\begin{equation} \label{eq:michel}
{\cal A} = 4 \frac{G_{\tau \ell}}{\sqrt{2}} \sum_{a = S,V,T}^{i,j = R,L} g_{ij}^a \langle \overline{\ell}_i | \Gamma^a| (\nu_{\ell})_n \rangle \langle (\overline{\nu}_{\tau})_m | \Gamma_a | \tau_j \rangle \, , 
\end{equation}
for $\ell =e,\mu$. Here $G_{\tau \ell}$ is the Fermi coupling singularized for each process, $\Gamma^a$ indicate the scalar, $\Gamma^S = 1$, vector $\Gamma^V = \gamma^{\mu}$ and tensor $\Gamma^T = \sigma^{\mu \nu}/\sqrt{2}$ interactions  and, finally, $L$ and $R$ the left- and right-handed chiralities of the electrically charged leptons, respectively. For a fixed set ${a,i,j}$ the neutrino chiralities $n$ and $m$ are also determined. The Standard Model predicts that
$g_{LL}^V = 1$ while all the rest are zero. 
The present situation is described by the figures in Table~\ref{tab:2}.  It can be seen that there is still room for improvement and the phenomenological analyses
of lepton decays of the tau lepton need to be pursued to settle our knowledge on the dynamics of the interaction. 
\end{multicols}
\begin{table}[h]
\begin{center}
\renewcommand{\arraystretch}{1.5}
\begin{tabular}{|c|c|c|c|}
\hline
$|g_{RR}^S| < 0.72$ & $|g_{LR}^S| < 0.95$ & $|g_{RL}^S| \leq 2$ & $|g_{LL}^S| \leq 2$ \\
\hline
$|g_{RR}^V| < 0.18$ & $|g_{LR}^V| < 0.12$ & $|g_{RL}^V| < 0.52$ & $|g_{LL}^V| \leq 1$ \\
\hline
$|g_{RR}^T| \equiv 0$ & $|g_{LR}^T| < 0.08$ & $|g_{RL}^T| < 0.51$ & $|g_{LL}^T| \equiv 0$ \\
\hline
\end{tabular}
\end{center}
\caption{\label{tab:2} Bounds on the Michel parameters from the decay $\tau^- \rightarrow \nu_{\tau} \mu^- \overline{\nu}_{\mu}$, at $95 \, \%$ CL \cite{ParticleDataGroup:2020ssz}. Notice that the tensor operators corresponding to equal chiralities vanish identically. We implement this information by setting vanishing couplings $g_{LL}^T$ and $g_{RR}^T$. } 
\end{table}
\begin{multicols}{2}

\subsection{Hadron decays}
\label{subsec:HSM}
The right-hand diagram in Fig.~\ref{fig:1} is the leading contribution to the process of production of hadrons through $\tau^- \rightarrow \nu_{\tau} (\overline{u}d, \overline{u} s)$. The amplitude for $\tau^- \rightarrow \nu_{\tau} H$, where $H$ is short for a hadron final state, is given by
\begin{equation} \label{eq:hadron}
{\cal A}  = \frac{G_F}{\sqrt{2}} V_{\mbox{\tiny CKM}} \, \overline{u}_{\nu_{\tau}} \gamma_{\mu} (1-\gamma_5) u_{\tau} \langle H| (V^{\mu} - A^{\mu}) e^{i L_{\mbox{\tiny QCD}}} |\Omega_{\mbox{\tiny H}} \rangle \, , 
\end{equation}
where $V_{\mbox{\tiny CKM}}$ is the relevant CKM matrix element and $\Omega_{\mbox{\tiny H}}$ is the hadron vacuum. Here, $V_{ij}^{\mu} = \overline{q}_j \gamma_{\mu} q_i$ and $A_{ij}^{\mu}=\overline{q}_j \gamma_{\mu} \gamma_5 q_i$ are the corresponding vector and axial-vector QCD currents, being $i,j= u,d,s$ the flavour indices. They will depend on the flavour content of the hadron final state $H$ \footnote{Another frequent notation for the QCD currents is $V_{\mu}^i = \overline{q} \gamma_{\mu} (\lambda^i/2) q$ and $A_{\mu}^i = \overline{q} \gamma_{\mu} \gamma_5 (\lambda^i/2) q$, being $q^T=(u,d,s)$ and $\lambda^i$, $i=1,...,8$, the Gell-mann matrices.}.
 In Eq.~(\ref{eq:hadron}) notice, in particular, the exponential of the QCD Lagrangian. It reminds us that the hadronization has to 
be carried out in the presence of the strong interaction. The determination of this matrix element is straightforward (feasible) for quarks in the final state, a 
process that gives relevant information on the {\em inclusive decays} of the tau lepton, i.e. in the sum of all hadron processes, but fails to convey the information of a particular decay channel with mesons in the final state, namely {\em exclusive decays}. There are several circumstances that explain this situation. Quarks are not observed in the final states and, therefore, a hadronization process has to be carried out to determine or paremeterize a particular decay. This implies the treatment of strong interactions at low energies, a regime where our knowledge of QCD is rather poor. The situation is even more involved because, with a mass a bit below 2 GeV, the tau lepton decays in an energy region populated by many hadron resonances and mesons.
\par 
In this Section we will briefly comment on both types of decays, their features, difficulties and the winnings we get from them. 
%
%
\subsubsection{Inclusive tau decays}
The analysis of the total tau hadron width, i.e. the sum of all meson final states in the decay of the tau lepton, reassures us the basics of QCD and it is able to gives us determinations of SM parameters \cite{Braaten:1991qm,Pich:1999hc,Gamiz:2002nu,Pich:2013lsa}. The relevant observable is the full width normalized to one of the leptonic decays, namely
\begin{equation} \label{eq:Rtau}
R_{\tau} \, \equiv \, \frac{\Gamma( \tau^- \rightarrow \nu_{\tau} \, \mbox{mesons})}{\Gamma(\tau^- \rightarrow \nu_{\tau} e^- \overline{\nu}_e )} \, ,
\end{equation}
where the radiation of final state photons in numerator and denominator are usually taken into account. 
It is customary to separate $R_{\tau} = R_{\tau,S=0} + R_{\tau,S=1}$, where $S$ indicates the strangeness of the final states. The non-strange component is determined experimentally into vector (even number of pions) and axial-vector (odd number of pions) parts, although 
they have also other non-strange contributions. The $R_{\tau,S=1}$ component has an odd number of kaons in the final state. 
\par 
As we did in Section~2, we can perform a naive and simple estimate of $R_{\tau}$ from its decomposition above: $R_{\tau} \simeq N_C |V_{ud}|^2 + N_C |V_{us}|^2 = N_C (1-|V_{ub}|^2) \simeq N_C$, where we have used the unitarity of the CKM matrix and the tiny value of $|V_{ub}|$. Experimentally, $R_{\tau}$ can be determined in two ways, either by calculating the numerator as the sum of all possible tau decays into mesons, or extracting from the total width the leptonic decays
\cite{HFLAV:2019otj}
\begin{eqnarray} \label{eq:exprt}
R_{\tau}^{\mbox{\tiny exp}} & = & \frac{\sum_i \Gamma_i(\tau^- \rightarrow \nu_{\tau} \, \mbox{mesons})}{\Gamma(\tau^- \rightarrow \nu_{\tau} e^- \overline{\nu}_e)} \, = \, 3.6355 (81)  \, , \nonumber \\
R_{\tau}^{\mbox{\tiny exp}} & = & \frac{1-B_e - B_{\mu}}{B_e} \, = \, 3.6370 (75) \, , 
\end{eqnarray} 
where $B_{\ell} = \Gamma(\tau^- \rightarrow \nu_{\tau} \ell^- \overline{\nu}_{\ell})/\Gamma_{\tau}$, $\ell=e,\mu$, being $\Gamma_{\tau}$ the total width. These correct our estimate above by a $20 \%$.
\par
\begin{figure}[t]
\centering
     \includegraphics[scale=0.60]{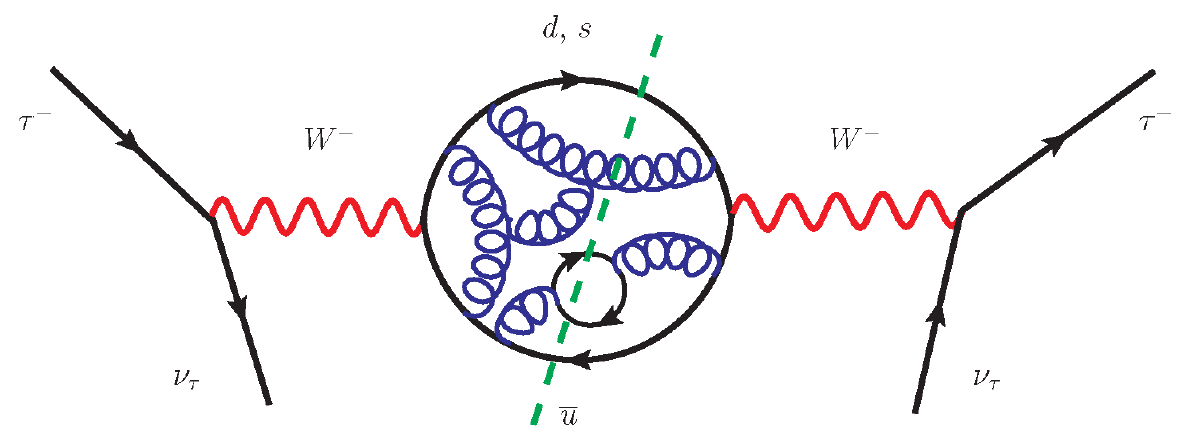}%
\hspace*{0.5cm}     \includegraphics[scale=0.65]{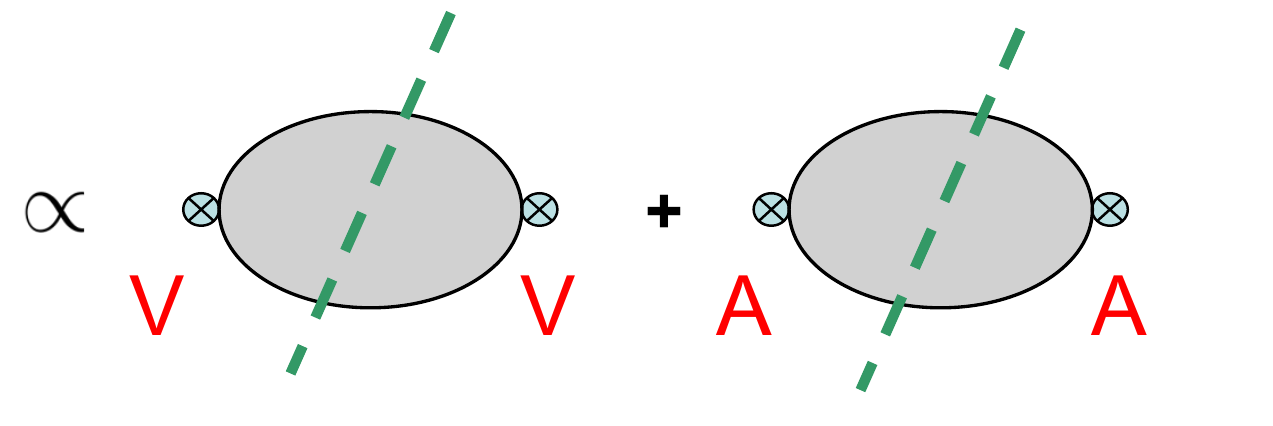}%
\caption{\label{fig:2} The full tau hadron decay width $\Gamma(\tau^- \rightarrow \nu_{\tau} \, \mbox{mesons})$ is proportional to the absorptive part of the hadron production, indicated here by the dashed line. This can be evaluated from the spectral decomposition (the imaginary part) of the VV and AA correlators. }
\end{figure}

 
We can give a more detailed account of the theoretical description of $R_{\tau}$. It can be shown that the hadron decay rate of the tau lepton can be written as an integral, over the invariant mass $s$ of the hadron final state, of the spectral functions \cite{Braaten:1991qm} (see Fig.~\ref{fig:2}),
\begin{eqnarray} \label{eq:rtauspectre}
R_{\tau} &=& 12 \pi \int_0^1 dx \left(1-x\right)^2 \left[ \left( 1+2x\right) \mbox{Im} \Pi^{(1)}(M_{\tau}^2 x)  \right.  \nonumber \\
&& \qquad \qquad \qquad \qquad \; \; \left. + \, \mbox{Im} \Pi^{(0)}(M_{\tau}^2 x) \right] \, ,
\end{eqnarray}
that correspond to
\begin{eqnarray} \label{eq:spectro}
\Pi^{(J)}(s) &\equiv & |V_{ud}|^2 \left( \Pi^{(J)}_{ud,V}(s)  + \Pi^{(J)}_{ud,A}(s)  \right) \Big|_{S=0}  \\
& & + \, |V_{us}|^2 \left( \Pi_{us,V}^{(J)} (s) + \Pi_{us,A}^{(J)}(s) \right)\Big|_{S=1} \, , \nonumber 
\end{eqnarray}
\end{multicols}
\begin{multicols}{2}
\noindent from the hadron correlators
\begin{eqnarray} \label{eq:correlr}
\Pi_{ij,{\cal J}}^{\mu \nu}(q)  & = & i \int d^4x e^{iqx} \langle \Omega_h | T {\cal J}_{ij}^{\mu}(x) {\cal J}_{ij}^{\nu}(x)^{\dagger} | \Omega_h\rangle \, \\
&= & (q^{\mu} q^{\nu} - q^2 g^{\mu \nu}) \Pi^{(1)}_{ij,{\cal J}}(q^2)  + q^{\mu} q^{\nu} \Pi^{(0)}_{ij,{\cal J}}(q^2) \, . \nonumber 
\end{eqnarray}
In these expressions $J$ indicates the total angular momentum of the hadron final state and ${\cal J}_{\mu}=V_{\mu},A_{\mu}$ the vector or axial-vector QCD currents.
\par
\begin{figure}[H]
\centering
   \hspace*{-0.4cm}  \includegraphics[scale=0.5]{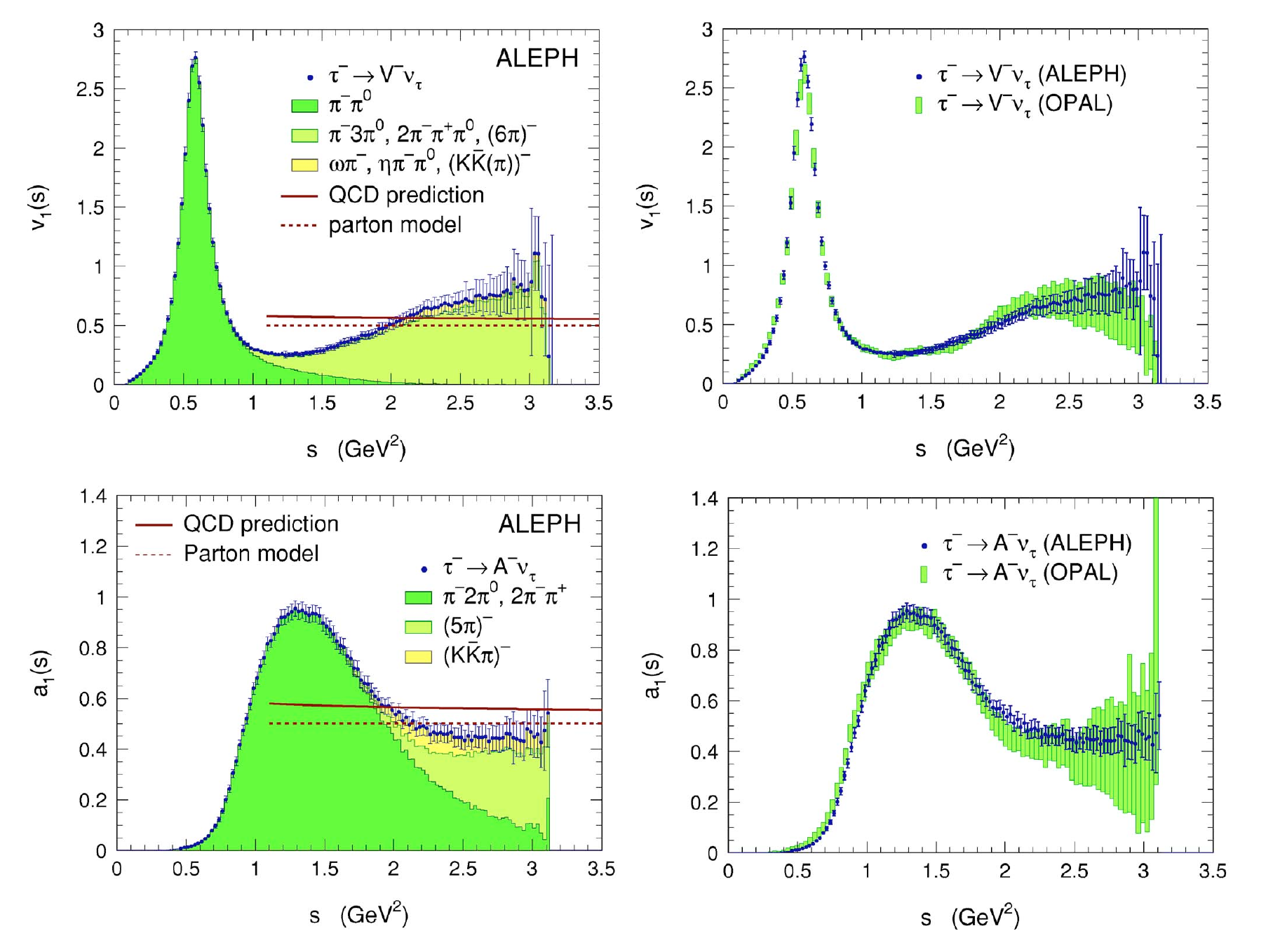}%
\caption{\label{fig:3} Experimental measurement of the $S=0$ vector and axial-vector spectral functions by the ALEPH and OPAL experiments at LEP II \cite{Davier:2005xq}.
Here $u_J = 2 \pi \, \mbox{Im} \Pi^{(J)}_{ud,U}$ for $u=\mbox{{\fontfamily{qcr}\selectfont v,a}}$ and $U=V,A$. In the vector function is clearly seen the contribution of the $\rho(770)$, while in the
axial-vector function it is clear the contribution of the $a_1(1260)$ and its prominent width.}
\end{figure}

 
The spectral decomposition, i.e. the imaginary part of these correlators, can be observed experimentally as the sum of all possible final states with mesons. I show
in Fig.~\ref{fig:3} the results by ALEPH and OPAL at LEP II  as collected in \cite{Davier:2005xq}.
\par 
By including those data in Eq.~(\ref{eq:rtauspectre}) and decomposing $R_{\tau,S=0}$ into its vector and axial-vector parts,  $R_{\tau,S=0}= R_{\tau,V} + R_{\tau,A}$,
it is obtained \cite{Davier:2005xq,Davier:2008sk}
\begin{eqnarray} \label{eq:rvasexp}
R_{\tau, V} &=& 1.783 (11)_{\mbox{\tiny exp}} (2)_{\mbox{\tiny V/A}} \, , \nonumber \\
R_{\tau, A} &=& 1.695 (11)_{\mbox{\tiny exp}} (2)_{\mbox{\tiny V/A}} \, , \nonumber \\
R_{\tau, S} &=& 0.1615 (40) \, .
\end{eqnarray}
Here the second error is due to a possible mishap in the identification of the vector or axial-vector contribution.
\par 
The phenomenological determination of the $R_{\tau}$ observable permits to obtain predictions for some SM observables as the strong coupling $\alpha_S(M_{\tau}^2)$
\cite{Braaten:1991qm}, the mass of the strange quark \cite{Pich:1999hc}, or the CKM matrix element $V_{us}$ \cite{Gamiz:2002nu} (see, for instance,\cite{Pich:2013lsa} 
for a detailed account). I will sketch the procedure in the case of the QCD strong coupling constant. 
\par 
The hadron correlators $\Pi^{(J)}(s)$ (\ref{eq:spectro}) are analytic everywhere in the complex plane, except on the positive real axis. Hence, we can use Cauchy's theorem to re-write the
expresion in Eq.~(\ref{eq:rtauspectre}) in terms of the full correlators, i.e.
\begin{eqnarray} \label{eq:rtauela}
R_{\tau} &=& 6 \pi i \oint_{|x| = 1} dx (1-x)^2 \left[ (1+2x) \Pi^{(0+1)}(M_{\tau}^2 x) \right. \nonumber \\
& & \qquad \qquad \qquad \qquad \; \; \; \;  \left. - 2x \, \Pi^{(0)}(M_{\tau}^2 x) \right] \, ,
\end{eqnarray} 
and the correlators can be parameterized by a dimensionally driven set of gauge-invariant scalar operators, an Operator Product Expansion (OPE), as
\begin{equation} \label{eq:ope}
\Pi^{(J)}(s) = \sum_{D=0,2,4...} \frac{1}{(-s)^{D/2}} \, \sum_{[{\cal O}] = E^D} C_D^{(J)}(s,\mu) \, \langle {\cal O}_D(\mu) \rangle \, .
\end{equation}
Notice the new $\mu$ parameter dependence. This is a new factorization scale which separates non-perturbative effects, hidden in the vacuum matrix elements
of the operators,  and short-distance physics in the Wilson coefficients. Obviously, the correlator does not have a $\mu$ dependence that, accordingly, has to 
cancel. The $D=0$ part corresponds to the unit operator and it gives the contribution
of perturbative QCD only, with massless quarks. Their masses enter in the $D=2$ term and $D=4$ already includes non-perturbative physics. 
%
%
\par
The fact that vector and axial-vector currents are conserved in the chiral limit, implies that only the $\Pi^{(0+1)}(s)$ correlator contributes in Eq.~(\ref{eq:rtauela}), and the polynomial part in front suggests that the $V+A$ is a clean observable because it has a dominant perturbative contribution, while non-perturbative effects arise at least at $D=6$.  We can write these contributions as
\begin{equation} \label{eq:rtaudeltas}
R_{\tau,{\mbox{\tiny V+A}}} = N_C |V_{ud}|^2 S_{\mbox{\tiny EW}} \left[ 1 + \delta_{\mbox{\tiny P}} + \delta_{\mbox{\tiny NP}} \right] \, , 
\end{equation}
where $S_{\mbox{\tiny EW}} = 1.0201(3)$ contains some electroweak corrections \cite{Marciano:1988vm,Erler:2002mv}. 
\par
The perturbative contribution is very sensitive
to $\alpha_S$ and  can be
written in terms of the $\alpha_S(\mu)$ coupling as $\delta_{\mbox{\tiny P}} = \sum_{n=1} K_n A^{(n)}(\alpha_S)$ 
\cite{LeDiberder:1992jjr,LeDiberder:1992zhd,Baikov:2008jh}, where the $K_n$ coefficients are known up to ${\cal O}(\alpha_S^4)$ and 
\begin{equation} \label{eq:ans}
A^{(n)}(\alpha_S) = \frac{1}{2 \pi i} \oint_{|x|=1} \frac{dx}{x} \left( \frac{\alpha_S(-M_{\tau}^2 x)}{\pi} \right)^n P(x) \,  ,
\end{equation}
with $P(x) = 1-2x+2x^3-x^4$. This function only depends on $a_{\tau} = \alpha_S(M_{\tau}^2)/\pi$ and the integrals are expanded in powers of this parameter. 
There is a well known incertitude in the evaluation of these integrals because the sizable value of the QCD coupling constant at the scale of $M_{\tau}$.
Hence, $R_{\tau}$ has a significant dependence on higher-order perturbative corrections. There are, essentially, two procedures that are usually used: i) an
expansion of $A^{(n)}(\alpha_S)$ in powers of $\alpha_S(M_{\tau}^2)$ and truncating the integrand to a fixed perturbative order in $\alpha_S(s)$ , called {\em fixed-order perturbation theory}, FOPT, and ii) using the exact solution for $\alpha_S(s)$ given by the renormalization-group $\beta-$function equation, called {\em contour-improved perturbation theory}, CIPT. See, for instance, Refs.~\cite{Pich:2016mgv,Pich:2016bdg,Boito:2016oam}. As a reference, the perturbative correction
in Eq.~(\ref{eq:rtaudeltas}) amounts $\delta_{\mbox{\tiny P}} \simeq 20 \%$. 
\par 
Let us now consider the non-perturbative correction $\delta_{\mbox{\tiny NP}}$ in Eq.~(\ref{eq:rtaudeltas}). This is parameterized by the power corrections in
Eq.~(\ref{eq:ope}) and it is given by
\begin{equation} \label{eq:nonp}
\delta_{\mbox{\tiny NP}} = \frac{-1}{2 \pi i} \oint_{|x|=1} dx (1-x)^2 (1+2x) \sum_{D \geq 2} \frac{ \overline{C_D} 
\langle {\cal O}_D(\mu)\rangle}{M_{\tau}^D (-x)^{D/2}}  \, , 
\end{equation}
with $\overline{C_D} \equiv C_D(M_{\tau}^2 x,\mu)$. In the case at hand, if we consider the chiral limit and neglect the $s$ dependence of the Wilson 
coefficients, the first contributing term in Eq.~(\ref{eq:nonp}) is the one with $D=6$ in the OPE expansion, i.e.
there is a suppresion factor $1/M_{\tau}^6$, at least, in the leading contribution to $\delta_{\mbox{\tiny NP}}$. The hadronic vacuum expectation values of
operators in Eq.~(\ref{eq:nonp}), for $D \geq 2$, are called {\em QCD condensates}. They parameterize the strong non-perturbative corrections and, in principle,
can be determined using lattice \cite{McNeile:2012xh}, phenomenology \cite{Jamin:2002ev} or QCD sum rules \cite{Shifman:1978bx,Shifman:1978by}, for instance.
The most updated analysis of ALEPH data \cite{Davier:2013sfa} gives $\delta_{\mbox{\tiny NP}} = -0.0064 (13)$, as expected much smaller than the 
perturbative correction and in good agreement with the theoretical prospect \cite{Braaten:1991qm}.
\par 
Hence, a determination of $\alpha_S(M_{\tau}^2)$ results from this procedure. They differ basically in the analyses carried out in the perturbative component
of $R_{\tau}$, as commented above. I quote some of the latest determinations:
\begin{eqnarray} \label{eq:detalphas}
\alpha_S(M_{\tau}^2) &=& 0.328 (13) ,  \qquad \mbox{\cite{Pich:2016bdg}} \nonumber \\
\alpha_S(M_{\tau}^2) &=& 0.308 (8) ,  \qquad  \; \;  \mbox{\cite{Boito:2020xli}} \nonumber \\
\alpha_S(M_{\tau}^2) &=& 0.312 (7) ,  \qquad  \; \; \mbox{\cite{Ayala:2021mwc}} .
\end{eqnarray}
They are in good agreement and their differences show the size of the incertitude in the determination of this parameter. 
\subsubsection{Exclusive tau decays}
\begin{figure}[H]
\centering
     \includegraphics[scale=0.46]{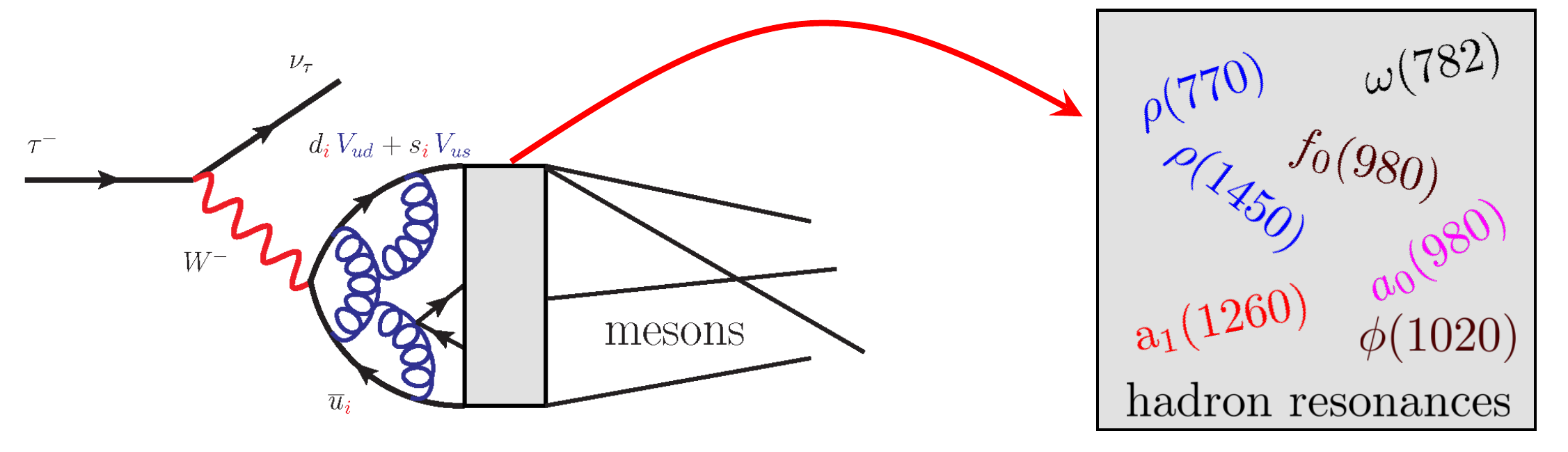}%
\caption{\label{fig:4} Form factors carry the information of the hadronization procedure in exclusive decays, in particular of the intermediate 
hadron resonances. }
\end{figure}

Let us consider now the study of decays of the tau lepton into specific hadron channels. We can come back to Eq.~(\ref{eq:hadron}) and ponder a particular
hadron channel $H$. This will have some possible quantum numbers (angular momentum, isospin, parity, ...) that we have to care about in our description. 
Hence, it is customary to parameterize the hadron matrix element as 
\begin{equation} \label{eq:hadronexclu}
 \langle H| (V_{\mu} - A_{\mu}) e^{i L_{\mbox{\tiny QCD}}} |\Omega_{\mbox{\tiny H}} \rangle  = \sum_i L^i_{\mu} \, F_i(Q^2,s,...) \, , 
\end{equation}
where $L^i_{\mu}$ indicates all possible Lorentz structures written with all the independent momenta of the process and respecting all known symmetries and
quantum numbers, and $F_i(Q^2,s....)$ are scalar functions of the independent invariants. The later are the {\em form factors} of the 
process \cite{Portoles:2007cx}. 
Form factors contain the information of the hadronization, Fig.~\ref{fig:4}, and their construction and determination provides the description of these decays.
Their theoretical construction belongs to the non-perturbative energy region of QCD and, in consequence, relies in models of the interaction. Successful  results 
come from phenomenological approaches based on Breit-Wigner descriptions of the resonances \cite{Kuhn:1990ad,Finkemeier:1995sr}, in the use of resonance chiral theory
\cite{Guerrero:1997ku,GomezDumm:2003ku,Dumm:2009va,Dumm:2009kj,Garces:2017jpz} or dispersion relations \cite{Pich:2001pj,GomezDumm:2013sib}.
\end{multicols}
\begin{multicols}{2}
Phenomenologically, it is known that form factors behave smoothly at high transfer of momenta \cite{Brodsky:1973kr,Lepage:1980fj}. 
This can be understood from the properties of the vector and axial-vector two-point correlators (\ref{eq:correlr}). 
They were studied, within perturbative QCD, in Ref.~\cite{Floratos:1978jb} where it was shown that both spectral functions
go to a constant value at infinite transfer of momenta, namely $\mbox{Im} \Pi_{ij,V/A}^{(1)} (q^2) \longrightarrow N_C/(12 \pi)$ as 
$q^2 \rightarrow \infty$, in the chiral limit and at one-loop in QCD. By local duality this can be understood as the sum of infinite positive contributions of intermediate hadron states, hence if the infinite sum gives a constant, heuristically it can be expected 
that any one of the contributions vanishes in that limit, and this behaviour translates into the form factors. 
\par 
A complementary framework used in the study of exclusive decays is the one provided by the {\em structure functions}, but we are not going to dwell on those here. See Ref.~\cite{Kuhn:1992nz} for a detailed explanation.
\par 
I will sketch some examples, namely the decays of the tau lepton into two and three pseudoscalars. The definition of form factors is, in general, not unique, but
the number of them for each process is fixed.  
\vspace*{0.2cm} \\
\noindent \underline{Two pseudoscalars} \\
The matrix element for the decay of the tau lepton into two pseudoscalars, $P=\pi,K,\eta,\eta'$, is driven by the vector current only. It has two form factors that can be defined as
\begin{eqnarray} \label{eq:ff2}
\langle P_1 P_2| V_{\mu} e^{i L_{\mbox{\tiny QCD}}} |\Omega_{\mbox{\tiny H}} \rangle & = & \left( g_{\mu \nu} -\frac{q_{\mu} q_{\nu}}{q^2} \right) t^{\nu} \, F_V(q^2)
\nonumber \\
& & + \,  q_{\mu} \, F_S(q^2) \, , 
\end{eqnarray}
where $q_{\mu} = (p_1+p_2)_{\mu}$ and $t_{\mu} = (p_1-p_2)_{\mu}$. Due to the conservation of the vector current in the $SU(3)$ limit, 
$\partial^{\mu} V_{\mu} \propto (m_i - m_j) \overline{q}_i q_j$,  the scalar form factor $F_S(q^2)$ only appears when the two pseudoscalars have different
masses, for instance in $\tau^- \rightarrow \nu_{\tau} K \pi$. Moreover, in $\tau^- \rightarrow \nu_{\tau} \pi^- \pi^0$ the contribution of the scalar form 
factor is tiny because it is an isospin breaking effect. 
\par 
The vector form factor of two pions, for instance, can be determined experimentally from the vector spectral function defined in Eq.~(\ref{eq:correlr}),
shown in $\mbox{{\fontfamily{qcr}\selectfont v}}_1(s)$ in Fig.~\ref{fig:3}.  They are related through $|F_V^{\pi \pi}(s)|^2 = 12 \, \mbox{{\fontfamily{qcr}\selectfont v}}_1^{\pi^- \pi^0}(s)/\beta^3$ with $\beta = \sqrt{1-4m_{\pi}^2/s}$. As can be seen in Fig.~\ref{fig:3} the dynamics of this form factor is dominated by the
contribucion of the $\rho(770)$ resonance. Hence, in the theoretical construction of the form factor we need to implement the role
of this resonance; moreover the asympotic behaviour of the form factor, commented above, gives that $F_V(q^2) \rightarrow 0$ for
$q^2 \rightarrow \infty$. All this information has to be included in the construction of this form factor. An efficient procedure
is the one designed by resonance chiral theory that, in addition, matches the chiral behaviour for $q^2 \ll M_{\rho}^2$, where 
$M_{\rho}$ is the mass of $\rho(770)$ \cite{Ecker:1988te,Ecker:1989yg,Guerrero:1997ku,Jamin:2006tk,Garces:2017jpz}.
\par
Notice that the vector current also drives hadronization in $e^+e^-$ scattering and, in consequence, the form factors of both processes are directly related. 
\vspace*{0.2cm} \\
\noindent \underline{Three pseudoscalars} \\
Both vector and axial-vector currents can contribute to this amplitude and, in the most general case, it is parameterized by four form factors:
$$ 
\langle P_1^- P_2^- P_3^+| (V_{\mu}-A_{\mu}) e^{i L_{\mbox{\tiny QCD}}} |\Omega_{\mbox{\tiny H}} \rangle = 
$$ 
\vspace*{-0.7cm}
\begin{eqnarray} \label{eq:ff3}
 \left( g_{\mu \nu} - \frac{Q_{\mu} Q_{\nu}}{Q^2} \right) \left[ t_{13}^{\nu} \, F_1^A(Q^2,s,t) +  t_{23}^{\nu} \,  F_2^A(Q^2,s,t) \right] & & \nonumber \\
 + \, Q_{\mu}\,  F_3^A(Q^2,s,t) + i \varepsilon_{\mu \alpha \beta \gamma} p_3^{\alpha} p_2^{\beta} p_1^{\gamma} \,  F_4^V(Q^2,s,t) \; , \; \; \;\; \; & & 
\end{eqnarray}  
where $Q_{\mu} = (p_1+p_2+p_3)_{\mu}$, $t_{ij}^{\mu} = (p_i-p_j)^{\mu}$, $s = (p_2+p_3)^2$ and $t=(p_1+p_3)^2$. The alphabetical label on the form factors indicate
the current that originates it, hence we have three axial-vector driven form factors and one coming from the vector current. 
\par
Each specific final state has its own
characeristics. For instance, the dominant channel is $\tau^- \rightarrow \nu_{\tau} \pi \pi \pi$ \cite{GomezDumm:2003ku,Dumm:2009va}, and has no contribution of the vector form factor in the isospin limit. Moreover the scalar $F_3^A(Q^2,s,t)$ is proportional to $m_{\pi}$ and then it vanishes in the chiral limit and, in any case, it gives a less important contribution compared with the other axial-vector form factors. Also in this channel, Bose symmetry requires that $F_1^A(Q^2,s,t) = F_2^A(Q^2,t,s)$. 
Phenomenological information on the axial-vector three-pion form factors can again be obtained from the experimental measurement of the axial-vector spectral function $\mbox{{\fontfamily{qcr}\selectfont a}}_1(s)$ in Fig.~\ref{fig:3}. This is the dominant
contribution of the spectral function and, it can be seen that it is dominated by the dynamics generated by the $a_1(1260)$ wide resonance.  What is measured in $\mbox{{\fontfamily{qcr}\selectfont a}}_1(s)$ is the partial width of the process as 
a function of $Q^2$ (the rest of kinematical variables have been integrated) and this is, naturally, a non-linear function of 
the axial-vector form factors $F_i^A(Q^2,s,t)$, for $i=1,2$. The precise relation is given, for instance, in Ref.~\cite{GomezDumm:2003ku}. The theoretical construction of these form factors relies, again, on the resonance dynamics and the asymptotic constraints
at high transfer of momenta. It has been carried out, for instance, within resonance chiral theory, in Refs.~\cite{GomezDumm:2003ku,Dumm:2009va}.
\par
Finally, all form factors contribute in the decay $\tau^- \rightarrow \nu_{\tau} K K \pi$ \cite{Dumm:2009kj}.
\section{Tau decays beyond the Standard Model}
\label{sec:BSM}
As was collected in Eq.~(\ref{eq:acci}) the Standard Model has a global symmetry that forbids the change of lepton flavour, or the number of leptons, in any process. As we already know that this symmetry is violated by neutrino mixing, there is no apparent reason
why processes with lepton flavour violation (LFV) in charged leptons should not occur, although still it has not been observed and the 
best upper-bound has been reached by the MEG experiment: $B(\mu^+ \rightarrow e^+ \gamma) < 4.2 \times 10^{-13}$ at $90 \%$ CL \cite{MEG:2016leq}.
\par
The search of lepton flavour or lepton number violations in processes with tau leptons, at present, cannot compete with muon related decays. However, as commented in the Introduction, the Belle-II experiment at SuperKEK will improve bounds, at least one order of magnitude, in tau decays. Belle-II has a specific program to look for LFV in decays, both hadron, i.e. $\tau \rightarrow \ell \pi$,
$\tau \rightarrow \ell \pi \pi$, $\tau \rightarrow \ell K$, etc., and lepton, i.e. $\tau \rightarrow \ell \gamma$, $\tau \rightarrow
\ell' \ell^+ \ell^-$ and so on, with $\ell, \ell' = e,\mu$. Their present bounds on those branching ratios lie between $10^{-7}$ and
$10^{-8}$ \cite{HFLAV:2019otj}. The bounds expected for Belle-II, with an estimated integrated luminosity of $50 \, \mbox{ab}^{-1}$, can be read from Ref.~\cite{Belle-II:2018jsg}, and are foreseen to lie around $B < 10^{-9} - 10^{-10}$.
\par 
SUSY \cite{Brignole:2004ah,Fukuyama:2005bh}  and $Z'$ \cite{Yue:2002ja} models, little Higgs \cite{delAguila:2011wk,Lami:2016vrs}, left-right symmetric models
\cite{Akeroyd:2006bb}, and others, have been applied in the analyses of LFV tau decays, giving branching ratios that lie
in the region at reach of the B-factories, i.e. ${\cal O}(10^{-7}-10^{-10})$. All these rely in the existence of a higher-energy scale, $\Lambda_{\mbox{\tiny LFV}} \gg \Lambda_{\mbox{\tiny EW}}$, being $\Lambda_{\mbox{\tiny EW}}$ the electroweak scale, such that the higher-dimensional non-renormalizable operators violating the lepton global symmetry (\ref{eq:acci}), arise. Based on this idea, a more model-independent framework is given by the Standard Model Effective Theory (SMEFT) at the electroweak scale \cite{Buchmuller:1985jz,Grzadkowski:2010es}, given by
\begin{equation} \label{eq:smeft0}
{\cal L}_{\mbox{\tiny SMEFT}} \, = \, {\cal L}_{\mbox{\tiny SM}} \, + \, \sum_{D > 4} \left( \frac{1}{\Lambda_{\mbox{\tiny LFV}}^{D-4}} \, \sum_i\, 
\, C_i^{(D)} \, {\cal O}_i^{(D)} \, \right) ,
\end{equation}
with ${\cal O}_i^{(D)}$ D-dimensional operators that contain the SM spectrum of particles and its fundamental symmetries, but breaking global lepton flavour
conservation, and $C_i^{(D)}$ are dimensionless Wilson coefficients determined by new physics. The lowest dimension operators giving LFV but conserving baryon number
have $D=6$. Analyses within this framework have been carried out \cite{Celis:2014asa,Husek:2020fru}. In the second reference we have studied several semihadron decays,
namely $\tau \rightarrow \ell P$, $\tau \rightarrow \ell PP$ and $\tau \rightarrow \ell V$, with $\ell = e,\mu$, and $P$ and $V$ pseudoscalar and vector mesons, respectively. In addition we have studied the lepton conversion processes $\ell N(A,Z) \rightarrow \tau X$, with $N(A,Z) = Fe(56,26)$ and $Pb(108,82)$, at the reach
of NA64 (CERN) \cite{Gninenko:2018num}. We have concluded that: i)  LFV tau decays constrain the dynamics stronger than the lepton conversion processes , though the 
later can be used to discern the relative weights of different contributing operators; ii) The Wilson coefficient $C_{\gamma}$ of the dipole operator 
${\cal O}_{\gamma} = \cos \theta_{\mbox{\tiny W}} {\cal O}_{\mbox{\tiny eB}} - \sin \theta_{\mbox{\tiny W}} {\cal O}_{\mbox{\tiny eW}}$ (notation of
Ref.~\cite{Grzadkowski:2010es})  happens to be the 
more constrained one, providing a foreseen result, from Belle-II, of $\Lambda_{\mbox{\tiny LFV}} > 330 \, \mbox{TeV}$ at $99.8 \%$ CL, for $C_{\gamma} = 1$. 
\par 
Finally, let us comment on lepton or baryon number violation. The remaining global symmetry in Eq.~(\ref{eq:acci}) indicates that total lepton, $L$, and baryon number,
$B$, are also conserved. These have a particular property, as the divergences of the corresponding currents are non-zero and equal. As a consequence they are 
anomalous, but $B-L$ is not. Because this anomaly, extensions of the SM cannot have a gauge boson enticing $B$ or $L$ violation, but it is possible to have one
driving $\Delta(B-L)=0$. Recently Belle published some results on 
$\Delta B=\Delta L=\pm1$ decays \cite{Belle:2020lfn}, for instance $\tau^- \rightarrow \overline{p} \mu^+ \mu^-$, which branching ratios bounded around $10^{-8}$. 
All these processes will also be a goal for Belle-II.
However, those branching ratios should be really tiny \cite{Fuentes-Martin:2014fxa}, because they should also provide channels of decay for the proton (with a virtual tau lepton), and we know that the lifetime of the proton  is huge.   
The analyses of these processes within SMEFT are eligible to present and future developments  \cite{Kobach:2016ami}.
For instance, it has 
been carried out an analysis with $D=5,7$ operators of the processes $\tau^+ \rightarrow \ell^- P^+ P^+$, with $\ell=e,\mu$ and $P$ a pseudoscalar meson 
\cite{Liao:2021qfj}. BaBar and Belle have looked for processes with $\Delta L=2$ but not involving baryons \cite{HFLAV:2019otj}.
\par 

\section{Conclusion}
\label{sec:conclu}
The physics of the tau lepton has many interesting aspects. The tau is the only known lepton that decays into hadrons, and this is reflected into a very rich, QCD driven, dynamics, both in the perturbative regime (inclusive processes) and in the non-perturbative energy region, for instance the study of hadronization of the QCD currents (exclusive processes).
\par 
The tau lepton offers a wide spectrum of processes in the study of violations of the SM global symmetries. The seeming violation of universality in B decays at LHCb or
the search for lepton flavour violation could disclose a new energy scale $\Lambda > \Lambda_{\mbox{\tiny EW}}$ where new physics lies. The Belle-II experiment will 
provide, in the next years, a large amount of information on tau decays, both for SM allowed processes and in the search of new physics. The theory has to be prepared to handle this future. 
\section*{Acknowledgements}
I would like to thank the organizers of the XIX Mexican School of Particles and Fields for their kind invitation. 
This work has been supported in part by MCIN/AEI/10.13039/501100011033 Grant No.~PID2020-114473GB-I00,
 by Grant No.\ MCIN/AEI/FPA2017-84445-P and
by PROMETEO/2017/053 and PROMETEO/2021/071 (Generalitat Valenciana).

\end{multicols}
\medline
\begin{multicols}{2}
%

\end{multicols}
\end{document}